# Environmental Effects on the Charge Transfer Properties of Graphene Quantum Dot Based Interfaces

Silvio Osella[1]*, Stefan Knippenberg[2,3]*

1. Chemical and Biological Systems Simulation Lab, Centre of New Technologies, University of Warsaw, Banacha 2C, 02-097 Warsaw, Poland.

2. Biomedical Research Institute, Hasselt University, Agoralaan Building C, 3590 Diepenbeek, Belgium.

3. Department of Theoretical Chemistry and Biology, School of Engineering Sciences in Chemistry, Biotechnology and Health, KTH Royal Institute of Technology, SE-10691 Stockholm, Sweden.

CORRESPONDING AUTHORS: s.osella@cent.uw.edu.pl; stefan.knippenberg@uhasselt.be

**Abstract**

Graphene quantum dots (GQD) are interesting materials due to the confined sizes which allow to exploit their optoelectronic properties, especially when they interface with organic molecules through physisorption. In particular, when interfaces are formed, charge transfer (CT) processes can occur, in which electrons can flow either from the GQD to the absorbed molecule, or *vice versa*. These processes are accessible by modelling and computational analysis. Yet, the presence of different environments can strongly affect the outcome of such simulations which, in turn, can lead to wrong results if not taken into account. In the present multiscale study, we assess the sensibility of the computational approach and compute the CT, calculated at interfaces composed by GQD and amino-acene derivatives. The hole transfer is strongly affected by dynamic disorder and the nature of the environment, and imposes stringent descriptions of the modelled systems to ensure enhanced accuracy of the transfer of charges.

**Introduction**

The rising of 2D materials have been boosted by the discovery of graphene and its derivatives, such as graphene nanoribbons (GNR) and graphene quantum dots (GQD), which present peculiar opto-electronic and transport properties due to the confinement of their size and the characteristic of being one-atom thick materials. In particular, GQD has been proven promising in application in the field of organic electronics due to their unusual properties; GQD are known to be superior in resistance to photobleaching, are low toxic and highly biocompatible.[1,2,3,4,5] Thanks to an abundance of raw material in nature, the cost is relatively low and production can be done on a large scale.[6,7,8,9] GQD emitters, which are sensitive to effects of structural and energetic disorder,[10] can be used in many opto-electronic applications such as bioimaging, photocatalysis, photovoltaics, lasers and light-emitting diodes.[11,12,13] Important are the possibilities to tune and enhance the magnetic as well as energy storage properties.[14] Another topic is the relationship between the chemical structure and optical properties of carbon nanodots: the excitation dependent emission originates from solvent relaxation effects of the solvents which are often used to suspend the carbon nanodots, while emission from dry samples does not show an excitation dependence.[15] To attenuate the solvent effect in the photo-luminous properties of the outgoing light, the carbon dot is often oxidized by nitric acid.[16] Nitrogen as an n-type dopant improves the conductivity, transport features and magnetic properties and is therefore used in spintronics and nanoelectronics.[17,18] In addition, they form stable interfaces with either donor or acceptor organic molecules.[19,20,21,22,23] Apart from the photochemical properties, the appeal of such interfaces is due to the strong charge transfer (CT) which can be generated by tuning the properties of the absorbed organic molecules. Yet, this CT can be strongly reduced by dynamic disorder (i.e. thermal fluctuations) and the presence of an environment which can strongly alter the transfer ability of the device, which is for instance found to be of importance for the electron transfer in biomolecular systems and DNA.[24,25,26,27] Hence, research has been focused on the design and synthesis of novel aromatic and conjugated organic materials which can efficiently transfer charges at the interface, to obtain high charge transfer ability.[28,29,30,31,32,33]

In this view, computations can play a central role in the prediction of the CT at the interfaces, since methods to calculate the transfer of charges are nowadays reliable and their prediction power can lead to a strong advantage for the scientific community thanks to a rational design of new interfaces, like it has been identified for discotic liquid crystals[34,35,36,37,38,39] Yet, a realistic model has to be taken into account, with inclusion of thermal fluctuations and the possible present of different environments, which might strongly alter the final CT ability of the interfaces. Moreover, a robust method for the quantification of the CT must be considered, such as the projective approach[40,41] which is of utmost importance when the interface is not formed by the same dimer molecules (i.e. is not a molecular crystal).

In this paper, the opto-electronic and charge transfer ability of interfaces of GQD and tetraaminoacenes (N-acenes) moieties are studied (Figure 1) by means of systematic multi-scale modelling techniques and (Time Dependent) Density Functional Theory [(TD)DFT] considering an effective environment. The N-acene moieties considered in this study are normally used as building blocks for longer or more complex acenes.[42,43,44] Yet, we observe that their particular chemistry –the presence of tetraamino groups- can be of potential interest for further application and have a function which requires a treatment which goes beyond a contribution as building block, such as transfer ability, which is usually not considered. Therefore, we investigated these acenes from N-naphtalene ('N-naphta') up to N-hexacene ('N-hexa').

**Methodology**

Density Functional Theory

The geometries of the N-acenes have been optimized at the Density Functional Theory (DFT) level considering the CAM-B3LYP long-range corrected functional.[45] Throughout all the study, Dunning's correlation consistent cc-pVDZ basis set has been used.[46] The usefulness of this approach for medium sized molecules has been proven in literature before.[47] It can be remarked that hybrid, long range corrected functionals are suited to describe local as well as non-local excitations, and that the asymptotic behavior is correct as the self-interaction issue has been solved, which was present in the older so-called GGA (Generalized Gradient Approximation) functionals. These functionals are therefore known to give rise to accurate band gaps. Although for the description of the geometries of these physisorbed compounds, in which van der Waals interactions are of importance, ab initio methods and even other functionals can be beneficial, too,[48] the choice for the functional in this work is based on the accuracy demands for not only the geometry and the description of the band gap, but also the excited state behavior, and the calculation of the charge transfer properties. A further validation of the method is explained in the following paragraph.

ADC calculations

To properly assess the optical properties of the N-acene molecules, an in-depth study of the excited state transitions has been performed by means of a high level correlated method, for which we selected the Algebraic Diagrammatic Construction scheme (ADC). [49,50] In order to enable a stringent comparison with the (TD)DFT results reported in the current study, the same CAM-B3LYP/cc-pVDZ geometries are used. In contrast to ADC(2)-s, which describes solely singly excited states, ADC(2)-x and ADC(3) report doubly excited states through first order in correlation. For these calculations, the cc-pVDZ basis set was used, too, and the analysis was performed by the Q-Chem 4.3 software.[51] In view of its computational cost and a scaling through $N^6$ with $N$ the number of basis functions, only shorter acenes (up to N-tetra) are analysed at the ADC(3) level of theory.

Electronic Coupling

Within the projective method,[40,41] the system is divided into fragments, in which an electron or hole is localized on a fragment and can hop from one fragment to another. The fragment calculations, performed by using the DFT method, follow a procedure where the orbitals of a pair of molecules (a dimer) are projected onto a basis set defined by the orbitals of each individual molecule (the fragments). The obtained set of orthogonal molecular orbital energies of the dimer are then used to express the Fock matrix in function of the new localized basis set, which enables the formation of a block-diagonal matrix. Here, we use a local version of a software which allows the study not only of vertical couplings (i.e. HOMO-HOMO, LUMO-LUMO) but also of diagonal contributions (i.e. HOMO-1-HOMO, HOMO-LUMO+1) which might be of importance when degeneration come into play. The system is thus split into two parts, a donor and an acceptor monomers and the coupling is then calculated considering any amount of molecular orbitals from the donor and acceptor sites, with all possible contributions (vertical and diagonal coupling terms). This software has been applied successfully in our previous studies on different interfaces.[52,53]

Molecular Dynamics

Molecular Dynamics calculations have been performed with the OPLS force field[54] within the GROMACS 2016.3 program.[55,56] Periodic boundary conditions were considered in 3 dimensions. Electrostatic interactions were treated by the particle-Mesh Ewald method[57] and bonds were constrained by the LINCS algorithm.[58] Electrostatics and van der Waals short-range interaction cutoffs were set to 1.6 nm. The NPT ensemble was used, with the Nosé−Hoover thermostat,[59,60] and a Parrinello−Rahman barostat[61] for a semi-isotropic pressure coupling at 1 bar and compressibility of $4.5 \cdot 10^{-5}$ bar$^{-1}$. RESP (restrained fit of electrostatic potential) partial atomic charges,[62] which were used in conjunction with the above called force field, were calculated at the CAM-B3LYP/ cc-pVDZ level of theory. Explicit water molecules were described by the TIP3P model.[63]

**Results and Discussions**

Being isolated (without the GQD) all these N-acenes have an energy gap in between 5-6 eV, which does not make them suitable for semiconducting applications (see Figure S1). Yet, once physisorbed on GQD, the energy gap remains constant along with the increased molecular length, at around 2.8 eV, making them semiconductor materials. This decrease of the energy gap at the interface is due to the different alignment of the frontier orbitals of the interface with respect to the orbitals of the separated fragments. In this view, both HOMO and LUMO orbitals of the interface are localized over the GQD, thus explaining the shrinkage of the gap (see Figure S2 for more details).

The CAM-B3LYP functional is not only used for the geometries of the N-acenes, but also for the analysis of their excited states, which are benchmarked against the ADC scheme of the second order in both strict (ADC(2)-s) and extended (ADC(2)-x) variants as well as the third order scheme (ADC(3)). The results of this benchmark are reported in Figure 1 (comparisons with ADC(2) and TDDFT/CAM-B3LYP results are provided in the Supplementary Information, Figures S3, S4 and Table S1).

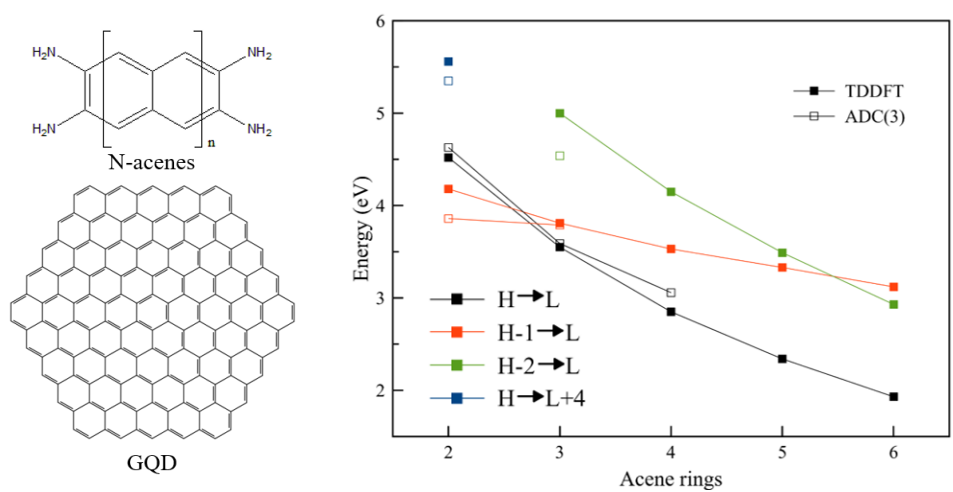

**Figure 1.** Chemical structures of the molecules forming the interfaces (left). The right plot reports the comparison between ADC(3) and TD-DFT/CAM-B3LYP for the low energy excited states transitions for the increasing N-acene lengths (without the GQD). The excited states are characterized by their main transitions, with *H* and *L* the highest occupied and lowest unoccupied molecular orbitals, respectively.

The lowest excited states of the parent acenes are known to yield an important double excitation character, which cannot be described using TDDFT and which requires therefore well-chosen *ab initio* methods. Famous is the case of the $2^1A_g$ and $1^1B_{2u}$ excited states; when the excitation energy is expressed in function of the number of fused acene rings, it is still far from trivial to simulate the correct order of both states, which might be reversed between pentacene and hexacene.[64] It is therefore of utmost importance to first assess the quality of a TDDFT level of theory for the here considered N-acenes. In fact, the frontier orbitals change their shape while increasing the molecular length, i.e. the HOMO level for N-naphta becomes HOMO-1 for longer N-acenes, and at the same time HOMO-1 becomes HOMO (see Figure S5). Thus, to use TDDFT to study the GQD/N-acene interfaces, it is essential to know that the description obtained is correct, and that it matches the one of the higher level methods. Figure 1 shows a quantitative agreement between the two methods, thus allowing the use of TDDFT for the study of the interfaces. In particular, both methods describe the low energy excited state as a HOMO to LUMO one, while the second is dominated by a HOMO-1 to LUMO transition. Moreover, maps of both CAM-B3LYP and Hartree-Fock orbitals used in ADC theory present an inverse order in energy for the HOMO and HOMO-1 going from N-naphta to N-antra. The third low energy transition has a more complicated character; the transition for N-naphta is from HOMO to LUMO+4 for both methods, but for longer N-acenes the agreement is less stringent. For these compounds with increased molecular backbones, TDDFT/CAM-B3LYP reports a steadily HOMO-2 to LUMO transition. For ADC(2) this assignment has only been found for N-penta, while for the other N-acenes the transition carries a HOMO to LUMO+2 character, yet high in energy. The same transition is found for the fourth excited state at the TDDFT/CAM-B3LYP level, which thus suggests that the use of the latter method can only be advised for low energy transitions.

Next, the GQD/N-acene interfaces were built and optimized at the CAM-B3LYP/cc-pVDZ level of theory.[45,46] As a practical model for the GQD and in line with work reported in the literature,[65] a $C_{150}H_{30}$ molecule has been considered. The final distance between the N-acenes and the GQD in our calculations has been found to be in between 3.54-3.58 Å. From our results, it follows that the optical properties of the N-acenes change when physisorbed on the GQD surface, as different transitions are observed for the low-lying excited states. In particular, the first transition has consistently a HOMO-1 to LUMO character from N-naphta to N-tetra, while for longer N-acenes the excited state is mainly characterized by a HOMO to LUMO transition. This reflects the different shape and localization of the frontier molecular orbitals (FMO) which change while increasing the molecular length (the shapes of the orbitals are reported in Figure S6). While for short N-acenes both HOMO and LUMO are localized over the GQD part of the interface, for N-penta and N-hexa the HOMO is localized over the N-acene backbone and the LUMO has mainly large contributions on the GQD. An interchange of occupied orbitals is seen going from N-tetra to N-penta, with the HOMO becoming HOMO-1. On the other hand, the HOMO-1 is stabilized and becomes HOMO-2. The HOMO-2 of N-tetra destabilizes when the backbone is enlarged; it can be identified as the HOMO of N-penta. Interestingly, this switch in orbitals is not present for virtual molecular orbitals. The different order of the FMOs is reflected in the different character of the transition from ground state to excited states. For interfaces up to GDQ/N-tetra, the first excited state transition has a localized character as the so-called $\Lambda$-parameter of Peach et al.[66] is larger than 0.4 (see Table 1). It is however seen that the $\Lambda$-value drastically drops to 0.1 for interfaces in which longer molecular backbones are considered, which is an indication for an excited state with a profound charge transfer character. The same situation is observed also for other transitions which involve the HOMO-2 (up to GQD/N-tetra) or HOMO (for interfaces with longer N-acenes), in which the orbital is localized over the N-acene moiety (see Table 1). A final word can here be said about the finite dimensions of the GQD, which is in general much bigger than the stacked N-acene molecule and which is also found to be rather in the center of the flake. For N-hexacene the practical boundaries have to be considered with care. However, the delocalization over the GQD, which is visible for the most important frontier orbitals in this study (HOMO-1, HOMO, LUMO, LUMO+1 – see Figure S6), guarantees that our results are free of spurious boundary effects.

**Table 1.** Low energy excited state transitions for the GQD/ N-acenes interfaces with increasing length calculated with the TD-DFT method. Excited states energies are in nm, the oscillator strength is reported in parentheses.

|       | TD-DFT        |      |            |
|-------|---------------|------|------------|
| ES=1  | Energy        | Λ    | transition |
| 2     | 1000 (0.003)  | 0.83 | H-1 -> L   |
| 3     | 998 (0.004)   | 0.86 | H-1 -> L   |
| 4     | 1000 (0.007)  | 0.84 | H-1 -> L   |
| 5     | 1034 (0.001)  | 0.11 | H -> L     |
| 6     | 1167 (0.00)   | 0.09 | H -> L     |
| ES=2  |               |      |            |
| 2     | 885 (0.026)   | 1.06 | H -> L     |
| 3     | 881 (0.03)    | 1.04 | H -> L     |
| 4     | 895 (0.008)   | 0.37 | H-2 -> L   |
| 5     | 998 (0.005)   | 0.84 | H-2 -> L   |
| 6     | 1076 (0.00)   | 0.08 | H -> L+1   |
| ES=3  |               |      |            |
| 2     | 733 (0.24)    | 0.24 | H-2 -> L   |
| 3     | 748 (0.002)   | 0.18 | H-2 -> L   |
| 4     | 880 (0.024)   | 0.81 | H -> L     |
| 5     | 966 (0.001)   | 0.11 | H -> L+1   |
| 6     | 998 (0.004)   | 0.82 | H-2 -> L   |
| ES=4  |               |      |            |
| 2     | 721 (0.61)    | 0.44 | H-2 -> L+1 |
| 3     | 718 (0.007)   | 0.18 | H-2 -> L+1 |
| 4     | 849 (0.002)   | 0.14 | H-2 -> L+1 |
| 5     | 885 (0.03)    | 0.95 | H-1 -> L   |
| 6     | 885 (0.028)   | 0.91 | H-1 -> L   |

This difference in shapes of the FMOs has a strong impact not only on optical properties, but also on charge transfer abilities and in particular on the strength of the hole coupling of the interfaces. To compute the charge transfer abilities of the interfaces the semi-classical Marcus theory in the non-adiabatic limit was considered:[67,68]

$$k(T) = \frac{2\pi}{\hbar} V^2 \frac{1}{\sqrt{4\pi\lambda k_B T}} exp\left[-\frac{\lambda}{4k_B T}\right] \qquad (1)$$

were $\lambda$ denotes the reorganization energy and $V$ the transfer integral between the initial and final states. For a fixed temperature, the large transfer rate can be attributed to the maximal transfer integral and the minimal reorganization energy. The transfer integral $V$ characterizes the strength of the electronic coupling between two adjacent molecules, which can be written as:[69]

$$V = \frac{J_{AB} - \frac{1}{2}(e_A + e_B)S}{1 - S^2} \qquad (2)$$

with $S$ being the overlap matrix and $e_A$ and $e_B$ the diagonal elements of a Hamiltonian defined by the frontier molecular orbitals of the isolated molecule A and B in the dimer representation, respectively. $J_{AB}$ refers to the coupling of either holes or electron, from molecule A to molecule B. Namely, for hole (electron) transfer, the HOMO and HOMO-1 (LUMO) should be considered, due to the (quasi)-degeneracy of the occupied energy levels. In an orthonormalized basis, these $e_n$ ($n= A, B$) give access to specific site energies $\varepsilon_n$ ($n= A, B$). The correction by means of these site energies is necessary in view of the pronounced polar character of the N-acene moieties and the different nature of both monomers in the complex. A variation of the projective method was used to quantify the transfer integral $V$ between the two non-equivalent components of the interface (more details in SI).[40,41] The calculated hole and electron transfer integral, as well as the site energies for the GQD/N-acene interfaces are reported in Figure 2.

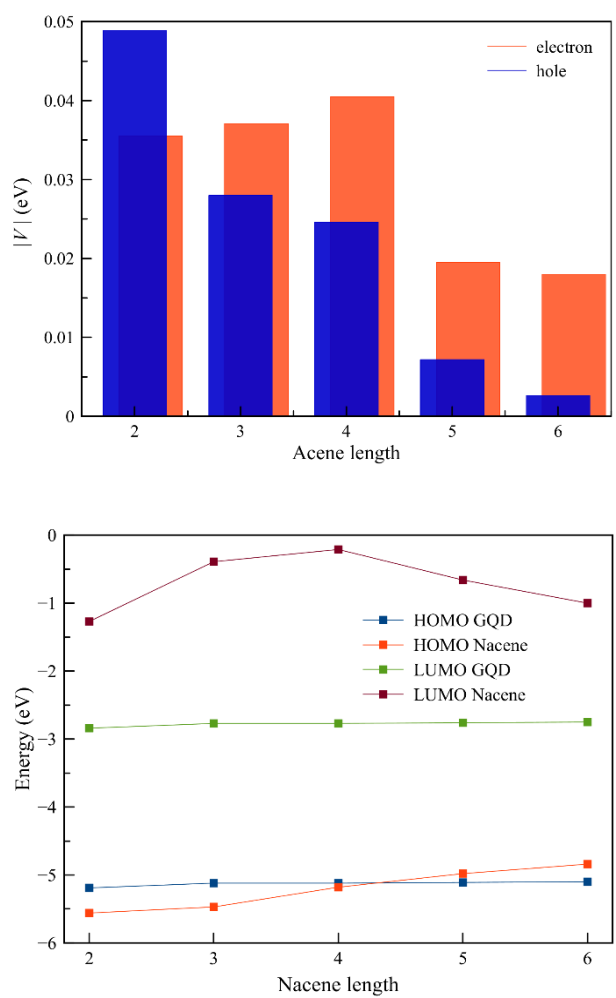

**Figure 2.** Electron and hole transfer integral (top) and site energies (bottom) values for the studied GQD/N-acene interfaces.

As a first result we observe that all the interfaces favour the transfer of electrons, with transfer integral values in between 20 and 40 meV. The values for the hole transfer integral decrease strongly with increasing the molecular length, and a range from 50 mV for N-naphta to 3 meV for N-hexa has been obtained. To rationalize the differences observed in the transfer integral values, the static disorder has to be considered, since it limits the mobility of the charge carriers.[70] The disorder is usually split into two contributions: the energetic disorder, accounting for the distribution of the site energies (corresponding to the distribution of the frontier level energies, i.e. HOMO and LUMO of the individual molecules) and the positional disorder, which describes the distribution and relative orientation of the molecules. The combination of the static disorder and the difference in orbital shapes of the interfaces with an increasing N-acene length, explains the difference in the calculated transfer integral. Since the site energies are highly sensitive to the respective orientation of the molecules,[71] a slight change in position and orientation can lead to large differences in the resulting transfer integral. For the interfaces studied here, we observe that the site energies (both HOMO and LUMO) of the GQD are stable. As expected for an increasing conjugated chain, however, pronounced differences in energy are noted when the length of the N-acenes is increased. Further, a drop in hole transfer is noted for the GQD/N-tetra and GQD/N-penta interfaces which can directly be related to the change in shapes of the highest occupied orbitals at the interface. The corresponding drop in electron transfer is more subtle, and should be related to the difference in site energies (different localization of the wavefunction) as well as differences in orientation of the N-acenes on the GQD.

In order to discuss a more realistic description of the interface, to take into account the dynamic disorder (thermal and geometrical fluctuations) present in a real case, as well as to account for the effect of an environment, the GQD/N-penta interface was selected and a 5 ns long classical molecular dynamics (MD) simulation was performed. The procedure used has been validated in a recent study of similar carbon nanodots (CDs) in different solvents.[72] Two MDs were conceived, the first with the interface embedded in vacuum, to account only for the dynamic disorder, and the second one in water to assess the effect of an environment in both fluctuations and transfer abilities. Although the normal acenes are insoluble in water, the N-acenes used in this study are soluble in acetic acid and other polar solvents. Thus, as first approximation of polar solvents, water was used. We would like to stress here that although the solvent might not be ideal for experimental purposes, the main aim of the study is to prove that solvent effects must be considered when computing the coupling. From these MD simulations, 25-30 independent snapshots were extracted in the last ns of simulation, and the transfer properties have been calculated (see Supporting Information for method details). The results, considering the degeneracy of the HOMO and HOMO-1 orbitals, are reported in Figure 3 and Table 2.

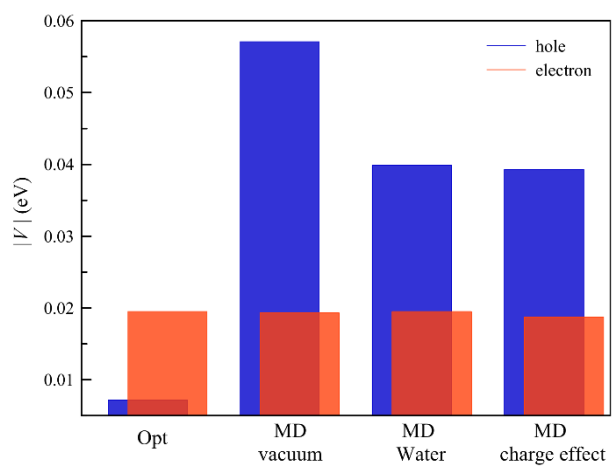

**Figure 3.** Electron and hole transfer integral values for the GQD/N-penta interface from Quantum Chemical optimizations and MD calculations. For the MD in vacuum, only dynamic disorder is considered; the same contribution is considered when a solvent is present in the MD ('MD water'). 'MD charge effects' refers to the transfer integral calculations in which the solvent molecules are considered as point charges, which can directly affect the coupling.

**Table 2.** Dependence of the calculated transfer integrals and site energies for the GQD/N-penta interface for hole (h) and electron (e) transfer on the different environment conditions. Note that the $V_{GQD/N-penta}$ values are given in meV, while $\varepsilon_{GQD}$ and $\varepsilon_{N-penta}$ are in eV.

| GQD/N-penta | carrier | $V_{GQD/N-penta}$ (meV) | $\varepsilon_{GQD}$ (eV) | $\varepsilon_{N-penta}$ (eV) |
|---|---|---|---|---|
| Opt | h | 7 | -5.11 | -4.98 |
|  | e | 20 | -2.76 | -0.66 |
| MD vacuum | h | 57 | -5.09 | -4.67 |
|  | e | 19 | -2.36 | -0.85 |
| MD water | h | 40 | -5.11 | -4.67 |
|  | e | 20 | -2.38 | -0.89 |
| MD charge | h | 39 | -5.27 | -4.67 |
|  | e | 19 | -2.62 | -0.89 |

We can observe that the electron transfer, with constant values around 20 meV, is not affected either by the introduction of thermal and configurational fluctuations nor the presence of the solvent as point charges in the transfer integral calculation. In contrast, a strong variation is present for the hole transfer in all these cases. The introduction of dynamic disorder to the optimized DFT structure leads to a strong enhancement of the hole transfer by one order of magnitude, from 7 to 57 meV. A less pronounced effect is observed when the MD calculations are performed with a water solvent, which enhances the hole transfer only up to 40 meV. Now, the GQD/N-penta interface has less freedom to move since the water environment stabilizes the interaction and hinders large reorientations and translations of the N-acene, which is located mainly on the centre of the GQD. The addition of point charges in the transfer integral calculation has a negligible effect on the hole transfer, which is enhanced up to 39 meV. Once more, the site energies should be considered to rationalize the observed behaviors. The HOMO and LUMO energies of the optimized GQD/N-penta interface are situated at -5.03 and -2.78 eV, respectively (see Figure S2). The site energies for the HOMOs are located at -5.11 and -4.98 eV for GQD and N-penta, respectively, and the LUMO ones amount to -2.76 and - 0.66 eV (Table 2). This picture changes when the site energies are considered for the snapshots extracted from the different MD simulations. An overall stabilization of the GQD HOMO is only observed when the charges of the water molecules are directly taken into account, while the N-penta HOMO (LUMO) site energy is destabilized (stabilized) by 30 (20) meV with respect to the quantum chemically optimized dimer. In addition, with the HOMO of the interface being localized over the N-penta fragment, the LUMO is localized over GQD. This rationalizes the stability of the electron transfer and the large variation observed for the hole transfer, which is strongly affected by the static disorder.

**Conclusions**

In summary, we described here the creation of novel GQD/N-acene interfaces in which the length of the N-acene backbone increases from N-naphta to N-hexa, and we calculated the transfer abilities of these interfaces by means of quantum chemical methods. We observed that the hole CT ability strongly decreases while increasing the N-acene length, while the electron CT abilities are more steady with values in between 20 and 40 meV. Considering the most commonly used acene (pentacene), we focus on the N-penta/GQD interface to study the effect of dynamic disorder (thermal and conformational fluctuations) and the presence on an explicit environment (liquid water) on the CT abilities of the selected interface. By means of the projective method, we assessed that both effects influence the change in CT abilities of the interface. In particular, we observed that hole transfer is strongly affected by dynamic disorder and the nature of the environment. It becomes even predominant with values in between 30 and 60 meV, while at the same time the electron transfer drops to 10-20 meV. Thus, neglecting both dynamic disorder and the environment lead to a wrong description of the CT abilities of the selected interfaces, going from an electron transfer to a hole transfer material. Therefore, efforts have to be done to both describe more realistic model systems and account for the different interactions exhibiting their influences in these environments, which are found to be indispensable to enhance the rational design of novel organic electronic materials.

**Acknowledgements**

S.O. is grateful to the Center for Quantum Materials and Nordita for his funding in Sweden. S.O. acknowledges the National Science Centre, Poland, grant UMO-2015/19/P/ST4/03636 for the funding from the European Union's Horizon 2020 research and innovation program under the Marie Skłodowska-Curie grant agreement No. 665778. The authors thank the Swedish Infrastructure Committee (SNIC) for the computational time granted within the medium allocations in 2016 (1-87, 1-415 and 1-465) and 2017 (1-16, 1-102), as well as the small ones 2015/4-44 and 2017/5-3. The calculations were partially performed at the Interdisciplinary Centre for Mathematical and Computational Modelling (ICM, University of Warsaw) under the G53-8 computational grant. The authors are also grateful to the Flemish Supercomputer Centre (VSC) and the Herculesstichting (Flanders, Belgium) for the calculation time on the Breniac cluster.

**Supporting Information**

Details of the benchmark, frontier orbitals, transfer integral analyses are reported.